# ON ELECTRON-HOLE SYMMETRY AND PHASE SEPARATION IN SOME ELECTRON DOPED CUPRATES

Lev P.Gor'kov,[a] Gregory B.Teitel'baum[b*]

[a]*NNHMFL, Florida State University, 1800 E P.Dirac Dr., Tallahassee FL 32310, USA*
[b]*E.K.Zavoiskii Institute for Technical Physics of the RAS, Sibirskii Trakt 10/7, Kazan 420029, RUSSIA*

**Abstract.**

We conclude from the analysis of the experimental NMR data for electron-doped cuprates that the Coulomb effects caused by doping lead to dynamical spatial phase separation that contributes to the nuclear spin relaxation. Remarkable, the "infinite-layer" $Sr_{0.9}La_{0.1}CuO_2$ reveals unexpected electron-hole symmetry. Its $^{63}Cu$ nuclear spin relaxation rate is the sum of a constant and the temperature dependent dissipation components, moreover, the latter turns out to be identical to the $1/^{63}T_1(T)$-behavior in the stoichiometric hole-type compound $YBa_2Cu_4O_8$. Connection to fluctuations of a magnetic sub-phase is discussed.

Soon after the discovery of hole-doped high-$T_c$ superconductor $La_{2-x}Ba_xCuO_4$ by Bednorz and Muller [1], Tokura and his coworkers synthesized an electron-doped superconductor $Nd_{2-x}Ce_xCuO_4$ [2]. The two systems are similar sharing the $CuO_2$-planes between the so-called block layers with rare-earth ions, and the insulating quasi two-dimensional antiferromagnetism at stoichiometric compositions. There are important differences, however, between the generic phase diagrams of the electron-doped and hole-doped materials. In order to elucidate the mechanism of high-$T_c$ superconductivity, it is very important to clarify the origin of similarities and differences between hole-doped (*p*-type) and electron-doped (*n*-type) cuprates.

The electron-doped High-Tc cuprates are not numerous, compared to their hole-doped analogs, possible, because of metallurgical reasons. Among them: 1) $Nd_{2-x}Ce_xCuO_{4-y}$ mentioned above [2] and similar materials with the another rare-earth ion instead of Nd; 2) so-called "infinite-layered" $Sr_{1-x}La_xCuO_2$ compounds that are mostly available as $Sr_{0.9}La_{0.1}CuO_2$ (see, e.g., in [3, 4]); 3) $Pr_{1-x}LaCe_xCuO_{4-y}$ (for references, see in [5]); and 4) a very interesting group $La_2$- $_x$RE$_x$CuO$_{4-y}$ (RE=Y, Lu, Sm, Eu, Gd, Tb); note the same valence of RE`s, i.e., the absence of an external doping [6]!

For both *n*-type and *p*-type the CuO$_2$-plane is considered be the playground, thus rising questions regarding common physics in two materials` groups. One should mention, though, that the electron-hole symmetry (see also discussion in [7]) is <u>far from being obvious, even improbable</u>. Indeed, stoichiometric (undoped) materials are always in the Mott insulating antiferromagnetic (AF) ground state. But transition to a metallic/superconducting (SC) state comes with doping. At hole doping (like in La$_{2-x}$Sr$_x$CuO$_4$) holes go initially on the oxygen sites. ARPES experiments [8] at low doping reveal so-called "Fermi surface arcs"; with further *x*-increase Cu $d^9$-levels hybridize with the oxygen band forming a large $(1+x)$-Fermi surface [9].

At electron doping added carriers expected to go directly on the Cu-sites, changing their electronic configuration ($d^9 \rightarrow d^{10}$) and inducing spin lattice dilution. Neutron experiments [10] for Nd$_{2-x}$Ce$_x$CuO$_{4-y}$ have shown that it remains magnetic and non-superconducting with *x* up to $x \sim 0.18$. The *instant* spin-spin correlations are antiferromagnetic and can be well described by the model of randomly-diluted localized one-half-spins with nearest-neighbors Heisenberg interaction. (The materials may be unstable and have an admixture of (NdCe)$_2$O$_3$ phase [11]).

Available ARPES data [12] show evolution with doping from the Mott-insulator Nd$_2$CuO$_4$ to a metal like small Fermi surface ($\sim x$) and finally to a large $(1+x)$-hole-like Fermi surface centered at $(\pi, \pi)$. Most striking fact, which certainly is *not* in favor of the *e-h*-symmetry in high temperature superconductors (HTSC), is the difference in the lattice structures: the *T*-symmetry for *p*-type *vs* the *T′*-symmetry in *n*-type [2]. As demonstrated, for instance, for La$_{2-x}$RE$_x$CuO$_4$ [6] the gradual *x*-decrease in these seemingly *metallic* materials with the *T′*-structure results in the sharp (1$^{st}$ order!?) transition (at $x \sim 0.05$) into an insulating state with the *T*-symmetry, suggesting that the Mott states for two symmetries may have different nature [6].

All the more unexpected, therefore, is the result obtained below that there <u>are</u> properties of the infinite-layers cuprates, Sr$_{0.9}$La$_{0.1}$CuO$_2$, that are <u>practically identical</u> to the ones of La$_{2-x}$Sr$_x$CuO$_4$.

There are not so many well-established facts about *n*-type cuprates yet. Thus, SC-order in them is not identified in all details. As to magnetic fluctuations, where the latter have been measured [13], they bear the *commensurate* AF character, unlike incommensurate, "stripe'-like fluctuations seen in the hole-doped materials [14]. Current experimental efforts are concentrated on attempts to establish more correspondence between other normal properties. At the issue are the existence of so-called "pseudogap" regime, phase separation/spatial inhomogeneities, search for "stripes" and study of magnetic properties in applied fields. These properties in the <u>hole-doped</u> cuprates strongly deviate from expectations based on the Fermi liquid theory and constitute the basic features of hole-doped cuprates in the main part of the (*T, x*)- phase diagram [15].

We concentrate our attention below on the NMR experiments for electron-doped cuprates, more specifically, on the data on the nuclear spin relaxation, $1/^{63}T_1$ [16, 17, 18]. Our point is that these are the only data so far that locally characterize the Cu-sites in a numerical fashion and, being less sample sensitive, as we believe, allow the quantitative comparison between the two types of cuprates. This approach has been successfully applied in [19] for unifying the properties of a broad group of *p*-type cuprates, and we start below with a brief discussion of some essential ideas.

At the very beginning of HTSC race it was suggested theoretically [20] that hole doped materials should manifest a tendency to inherent phase segregation involving both lattice and electronic degrees of freedom leading to a 1$^{st}$ order transition, which, however, remains frustrated dynamically due to Coulomb effects in the presence of embedded ionic charges. The tendency was proved in numerous experiments. Key role in understanding the universality of this phenomenon and its relevance to the HTSC mechanism belonged to neutron experiments. One of the most important results for the hole-doped materials was the observation of incommensurate (IC) AF fluctuations possessing some "stripe" structure [21]. The neutron scattering data are available only for the restricted number of electron doped superconductors with the $T'$ structure, such as $Nd_{2-x}Ce_xCuO_4$. As mentioned above, here AF fluctuations bear commensurate character.

In our paper [19] we addressed the phase diagram of hole-doped cuprates at temperatures above $T_c$, more specifically, inside so-called "pseudogap regime". It is known that many properties of hole-doped cuprates manifest a characteristic crossover at some $T^*(x)$ called in the literature by "pseudogap temperature" [15]. Specifically, the common practice for the NMR data to define $T^*$ [22, 23] was through the position of a maximum in the product $1/^{63}T_1T$.

Analysis [19] has shown that this approach is misleading. In a temperature interval above $T_c$ and below some temperature $T^*(x)$ for a broad class of cuprates with hole conductivity the pseudogap regime in this part of the phase diagram actually signifies a <u>two phase coexistence</u>. As the result, $1/^{63}T_1(x,T)$ itself can be decomposed into two independent dissipation mechanisms:

$$1/^{63}T_1 = 1/^{63}\overline{T}_1(x) + 1/^{63}\widetilde{T}_1(T) \qquad (1)$$

where the first terms comes about due to relaxation on the "stripe"-like dynamical excitations (that surprisingly leads to a temperature independent contribution!), and an ``universal" temperature dependent term. The "stripes"-term comes about only with external doping. Our estimations for $La_{1-x}Sr_xCuO_4$ [19] were in a quantitative agreement with the inelastic neutron scattering data for this compound [24].

The second, temperature dependent contribution, as it turned out, coincides with $1/^{63}T_1(T)$ measured for the defect-free stoichiometric material, $YBa_2Cu_4O_8$ (YBCO 124).

$T^*(x)$ now acquires the meaning [19] of the temperature for the onset of a 1st order phase transition frustrated by charge electroneutrality condition (the onset, $T^*(x)$, naturally is also seen as a crossover in the $T$-dependence of other physical characteristics, such as resistivity or the Hall coefficient).

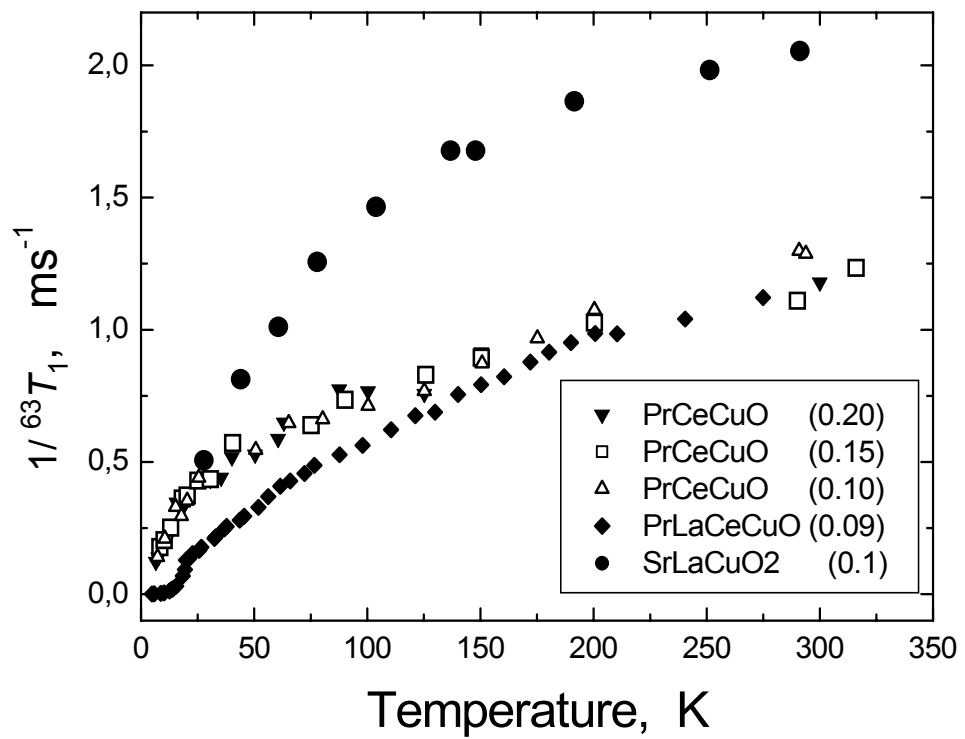

Fig. 1. The temperature dependence of the $^{63}$Cu nuclear spin relaxation rate for $Pr_{2-x}Ce_x CuO_4$ (for $x$=0.10; 0.15; 0.20  $T_c$=16K; 20K; and 20K correspondingly) [17], $Sr_{0.9}La_{0.1}CuO_2$ ($T_c$=43K) [17], $Pr_{0.91}LaCe_{0.09}CuO_4$ ($T_c$=26K) [18]. The doping level, $x$, is shown in the brackets.

With these ideas in mind we start with plotting in Fig. 1 the NMR data, $1/^{63}T_1$, for the electron-doped materials. The data exist only for the three groups of electron doped HTSC materials out of four mentioned in the introductory part of the paper. The first one includes the materials $(RE)_{2-x}Ce_x CuO_4$ with so called $T'$ structure (RE = Nd, Sm, Pr...) [25]. To avoid the additional contribution to the nuclear spin relaxation due to rare-earth magnetic moments we restrict ourselves with the case of nonmagnetic RE ions such as Pr or La. The second class of $n$-type cuprates is $(Sr, L)CuO_2$ ($L$= La, Sm, Nd, Gd…) with a so-called "infinite-layer" structure [3,4]. These materials have the simplest crystal structure among all HTSC`s consisting of an infinite stacking of $CuO_2$ planes and (Sr, L) layers with no charge reservoir block commonly presenting in cuprates. They also have the stoichiometric oxygen content without vacancies or interstitial oxygen, so the possible disorder comes to play only due to the doping with the rare-earth ions [26]. Finally, for completeness, we have also plotted data for $Pr_{1-x}LaCe_xCuO_4$ [18]; here $T_1T = const$ in normal state ($H > H_{cr}$) but the slope starts to vary above 70K.

For the $Pr_{2-x}Ce_xCuO_4$ and $Pr_{1-x}LaCe_xCuO_4$ materials, the $1/^{63}T_1$ behavior looks somewhat irregular, and we will return to it later. At the same time the T-dependence of the relaxation rate for the Cu-nuclear spins in the "infinite-layered" $Sr_{0.9}La_{0.1}CuO_2$ immediately attracts the attention by its resemblance to the corresponding data for $La_{1-x}Sr_xCuO_4$ ( see Fig. 1 in [19] and references therein), and we attempted to repeat the decomposition of Eq. (1) also for this case. Bluntly speaking, we applied a *vertical* offset to the $Sr_{0.9}La_{0.1}CuO_2$ data to see whether it is possible to superimpose the curve on the T-dependence of $1/^{63}T_1(T)$ for YBCO 124 [27, 28].

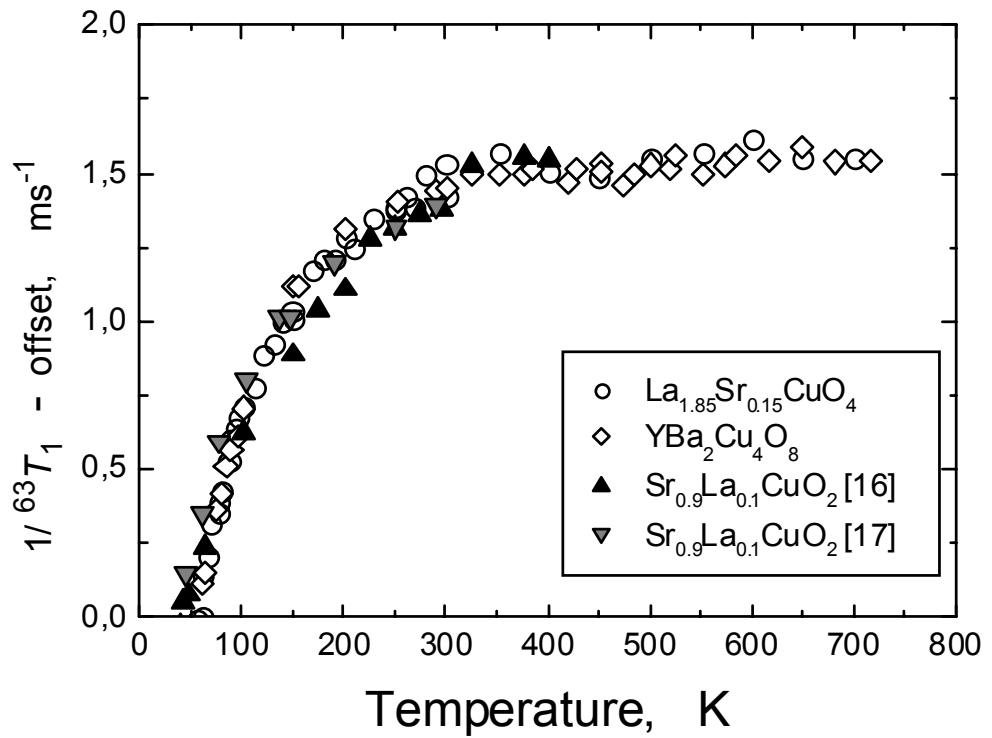

Fig. 2. The temperature dependence of the $^{63}Cu$ nuclear spin relaxation rate for different h- and n-doped materials after the corresponding vertical offset: $YBa_2Cu_4O_8$ (offset = 0), $Sr_{0.9}La_{0.1}CuO_2$ (offset = 0.64), $La_{1.85}Sr_{0.15}CuO_4$ (offset = 1.4).

The result is seen in Fig. 2 - the agreement, indeed, is very good (above $T_c$=82K for YBCO124). The value of the offset, i.e., the corresponding T-independent term in Eq.(1) equals: $1/^{63}\overline{T}_1(x) = 0.64$ ms$^{-1}$.

This is the main result of the paper. The temperature dependence in Fig. 1, according to fit in Fig. 2, just coincides with the one for *hole doped* stoichiometric YBCO 124. Note that the *T*-independent component is *positive*, i.e., presents an independent contribution to the dissipation mechanisms.

As it was already mentioned above, the $1/^{63}\overline{T}_1$ in hole doped materials were nonzero [19] for those of them for which incommensurate (IC) peaks have been observed in neutron scattering [14]. Peaks are close to the [$\pi, \pi$] – point: at [$\pi(1\pm\delta), \pi$] and [$\pi, \pi(1\pm\delta)$] [15] and are commonly treated as manifestation of the dynamical stripe fluctuations [29].

We suggest that nonzero "offset" obtained for the electron doped superconductor $Sr_{0.9}La_{0.1}CuO_2$ may also be the indication of the phase separation in form of stripes. Of course one cannot exclude that another kind of the dynamical phase separation may also induce the temperature independent contribution to the relaxation rate.

The additional evidence in favor of the microscopic phase separation in $Sr_{0.9}La_{0.1}CuO_2$ was reported in [30], where the authors observed a considerable wipeout of the NMR signal starting at 200K. Since the part of nuclei remains NMR-active down to 4.2K it gives strong hints in favor of a broad distribution of the fluctuation frequencies in this compound. In its turn this indicates that the magnetic state is strongly inhomogeneous. The typical spatial scale of these inhomogeneities cannot be estimated from the NMR experiments. It is important, therefore, that the AF-ordered sub-phase has been seen in the recent *elastic* neutron scattering experiments in $Pr_{1-x}LaCe_xCuO_4$ even in the superconducting state [31]. This gives additional evidence in favor of coexistence of frozen AF order with superconductivity in *all n*-type HTSC (Coexistence of the field-induced AF-order with superconductivity follows also from the recent μSR data [7]).

Let us finally discuss in brief the NMR-data for the materials with the $T'$-structure. In Fig. 1 the temperature dependence of $1/^{63}T_1$ for $Pr_{2-x}Ce_xCuO_4$, was measured [17] for the Ce content, *x*, equal to 0.1; 0;15; and 0.20. One can see that in this case the temperature dependence differs considerably from that shown in Fig. 2 and cannot be described by Eq. (1). The relaxation rate $1/^{63}T_1$ is smaller and is practically independent on Ce doping. Nevertheless, above $T_c$ (depending on the Ce-concentration) the data are closer to a linear slope in *T*, reminding the Korringa behavior that, if extrapolated to lowest temperatures, would again give a constant term as in Eq. (1). In case of $Pr_{1-x}LaCe_xCuO_4$ at lower temperature, especially in the normal state in large field below $T_c$, the system demonstrates the metallic Korringa behavior; however, its slope also varies then at higher temperatures. Note that at the same time, the material demonstrates such Fermi liquid features as the $T^2$-dependence in resistivity [18]. We think that at the current stage the only reliable conclusion of a general character that one might draw from such data is that these systems should be rather inhomogeneous. Indeed, even for the AF material $Nd_{1-}$

$_x$Ce$_x$CuO$_4$ [2] it is known that itinerant carriers are present in the system [32, 33]. This inhomogeneity is expected: as it was noticed first in [20], the necessity to balance imbedded charges of the dopants` ions would restrict the growth of AF sub-phases. It is possible that better screening due to the presence of metallic carriers leads to smaller sizes of inhomogeneitis and to their dynamical character [20], thus providing for an additional mechanism in the NMR relaxation. (The *n*-type HTSC materials may have somewhat larger degree of disorder to pin fluctuations at low temperatures). It is worth to emphasize once again that in the neutron scattering studies of similar compounds [10, 31, 34, 35] only well-defined *commensurate* spin fluctuations were observed in both AF and SC phases, analogous to the results for Nd$_{1-x}$Ce$_x$CuO$_4$, unlike the ones for the hole-doped system.

To summarize, we commented, first, that most of known electron doped cuprates, according to the nuclear spin relaxation data, exhibit a phase separation, of a dynamical character above $T_c$, as confirmed by observation of (commensurate) AF fluctuations. The last fact, probably, is somehow related to the $T'$-symmetry of the most of *n*-doped materials.

Of more significance is our result above that the "infinite-layer" Sr$_{0.9}$La$_{0.1}$CuO$_2$ demonstrates the NMR –behavior such that not only reveal the electron-hole symmetry for this type of cuprates, but also shares quantitatively the temperature dependence of $1/^{63}T_1(T)$ with the broad class of the hole doped cuprates [19]. The result calls for careful inelastic neutron scattering studies of the AF fluctuations in this compound in a broad temperature range to clarify the physical reasons behind the decomposition of Eq. (1) discovered for this material.

The authors are grateful to J. Haase for valuable discussions and the consultations concerning the relevant literature. The work of L.P.G. was supported by the NHMFL through NSF cooperative agreement DMR-9527035 and the State of Florida, that of G.B.T. through the RFBR Grant N 04-02-17137.